\documentstyle[twocolumn,epsf]{jpsj}
\newcommand{\eps}{\varepsilon}
\newcommand{\vphi}{\varphi}

\newcommand{\til}[1]{\tilde{#1}}
\def\runtitle{Dimensional Crossover by a Local Inhomogeneity
in Soliton-Pair Nucleation}

\def\runauthor{Takeo {\sc Kato}}
\title{Dimensional Crossover by a Local Inhomogeneity \\
in Soliton-Pair Nucleation
\footnote{Submitted to J. Phys. Soc. Jpn.} 
}

\author{Takeo {\sc Kato}}

\inst
{Institute for Solid State Physics, University of Tokyo,
\\ 7-22-1 Roppongi, Minato-ku, Tokyo 106-8666}

\recdate{}

\abst{Soliton-pair nucleation rates $\Gamma = A \exp(-B)$
are studied in highly-biased sine-Gordon systems with
a local inhomogeneity for both a thermal activation regime and
a quantum tunneling regime. It is found that the local 
inhomogeneity strongly affects the nucleation rates by 
modifying the bias-dependence of the exponent $B$.
This change in the exponent $B$ is explained as a
dimensional crossover caused by the local inhomogeneity.
It is also shown that there is another crossover, at which
$A$ becomes independent of the system size.}

\kword
{Sine-Gordon equation, soliton-pair nucleation, 
local inhomogeneity, long Josephson junctions}

\begin{document}
\sloppy
\maketitle
%
%

The dynamics of a one-dimensional sine-Gordon (SG) model has
provided conventional understandings of various systems
in physics, including dislocation in crystals, charge density waves
(CDWs) in quasi-one-dimensional materials, and long Josephson
junctions. One of typical processes
in these systems is nucleation of soliton-antisoliton
pairs driven by an external force. The nucleation 
rates of soliton-pairs 
have been calculated in both a thermal activation regime
and a quantum tunneling regime for
homogeneous systems.~\cite{Hida84,Nakaya86,Ivlev87,Haenggi90a} 
Inhomogeneities, however, are unavoidable in actual
experimental situations. They may change the nucleation process
drastically, because the nucleation rate 
$\Gamma = A e^{-B}$ is very sensitive
to local modulation in the exponent $B$, which is proportional to
a barrier height. Especially, inhomogeneities may
change the bias-dependence of the exponent $B$, which can
be detected experimentally. This possibility has been
pointed out first in the context of CDW systems.~\cite{Yumoto00}

In this Letter, effects of a local inhomogeneity on
soliton-pair nucleation are studied at a high driving bias.
The nucleation rates are evaluated by Langer's
method.~\cite{Langer67} The bias-dependence of $B$ is typically
expressed as $B \propto (f_{\rm c}-f)^{\gamma}$, where $f$ and $f_{\rm c}$
are an external bias and a classical threshold bias, respectively.
It is shown that the inhomogeneity changes not only the threshold current
$f_{\rm c}$, but also the exponent $\gamma$. This change of $\gamma$ 
affects the nucleation rates drastically, and is expected to be detected
experimentally. It is claimed that the change of $\gamma$ is essentially
understood by a {\it dimensional crossover} caused
by the local inhomogeneity. Another crossover about the prefactor
$A$ is also discussed. Although the following discussion is
available for various physical systems described by the SG model,
long Josephson junctions (LJJs)
are considered as a comprehensive example, which allows
well-controlled experiments.

The classical equation of motion of LJJs with a local impurity
is given as~\cite{McLaughlin78}
\begin{equation}
\phi_{tt} - \phi_{xx} + \sin \phi - f - \eps \delta(x)
\sin \phi = 0,
\label{CLE}
\end{equation}
where $f$ is an external current density normalized by the
critical current density, and $\eps (>0)$ is a strength of an impurity
potential made by modifying the thickness of insulator layers locally.
Here, the spatial and temporal variables are normalized by the
Josephson length and plasma frequency in LJJs. In this Letter,
dissipation due to quasi-particle currents is assumed to be 
small, but not extremely weak, to guarantee the thermal 
equilibrium in a metastable state. The partition function
of this system is described by the imaginary-time path integrals as
$Z = \int {\cal D}\phi(x,\tau) \exp(- S_{\rm E}/g^2)$, where
\begin{eqnarray}
S_{\rm E}[\phi(x,\tau)] = \int^{L/2}_{-L/2} {\rm d}x \int^{g^2/T}_{0} 
\! {\rm d}
\tau \Biggl[ \frac{\phi_x^2}{2}+\frac{\phi_{\tau}^2}{2} 
\Biggr. \nonumber \\ \Biggl.
- f \phi + (1-\eps \delta(x) )(1 - \cos \phi ) \Biggr],
\label{SE}
\end{eqnarray}
is the Euclidean action and $g^2$ is the normalized Planck constant.
Here, $L$ and $T$ are the length of the junction and the temperature
normalized by the Josephson energy per unit length, respectively.
At high biases $f = 1-\eta$ ($\eta \ll 1$), the
potential energy is allowed to be expanded to a
quadratic-plus-cubic form. By changing the field variable
as $\phi(x)=\pi/2 + \sqrt{2\eta} (\vphi(x) - 1)$, and by
rescaling the spatial (temporal) variables as $x = (2\eta)^{-1/4} x'$
($\tau = (2\eta)^{-1/4}\tau'$), the Euclidean action is modified as
\begin{eqnarray}
S_{\rm E}[\vphi(x',\tau')] &=& (2\eta)^1 
\int {\rm d}x' \int {\rm d}\tau' 
\left[ \frac{\vphi_{x'}^2}{2} + \frac{\vphi_{\tau'}^2}{2} 
\right. \nonumber \\
&+& \left. \frac{\vphi^2}{2} - \frac{\vphi^3}{6}
- \til{\eps}\delta(x') \vphi \right].
\label{SE2}
\end{eqnarray}
Here, $\til{\eps} = \eps (2\eta)^{-3/4}$ is an effective impurity 
strength, which determines the magnitude of the inhomogeneity effects
on nucleation. It should be noted that the impurity effects
can be controlled by the external current $f=1-\eta$;
the effective impurity strength $\til{\eps} = \eps (2\eta)^{-3/4}$
is enhanced if the current approaches
the classical threshold current as $\eta \rightarrow 0$.

The nucleation rates are evaluated by Langer's
method in terms of the imaginary part of the free energy 
$F = -T \ln Z$.~\cite{Langer67,Callan77,Affleck81,Weiss93}
The partition function $Z$ is evaluated by integrating out 
the field $\vphi(x',\tau')$ up to the second order of
fluctuations around stationary solutions determined by
$\delta S_{\rm E}/\delta \vphi = 0$. In the present case,
there are two stationary solutions: one is a stable solution 
$\vphi_0(x',\tau')$ and the other one is a bounce solution 
$\vphi_{\rm B}(x',\tau')$ with one unstable mode. 
The evaluated partition function $Z$ includes an imaginary part
produced by the integration around the bounce solution
$\vphi_{\rm B}(x')$. The nucleation rate is then related to
the free energy through $\Gamma = 2 f(T) {\rm Im} F$.
Here, $f(T)$ is a temperature-dependent factor,
and takes $1$ for $T < T_0$, and $T_0/T$ for 
$T > T_0$, where $T_0$ is the crossover temperature between
the thermal activation regime and the quantum tunneling
regime.~\cite{Affleck81,Weiss93,Grabert84,Larkin84}
The nucleation rate $\Gamma = A\exp(-B)$ is then obtained as
\begin{eqnarray}
A &=& \frac{f(T) T}{g^2} \prod_{i=1}^{\infty} 
\left(\frac{\lambda_i^{(0)}}{\lambda_i^{({\rm B})}}\right)^{1/2}, 
\label{prefactor} \\
B &=& (S[\vphi_{\rm B}(x',\tau')]-S[\vphi_0(x',\tau')])/g^2.
\end{eqnarray}
Here, $\lambda_i^{(0)}$s ($\lambda_i^{({\rm B})}$s) 
are the frequencies of eigenmodes around $\vphi_{0}$ ($\vphi_{\rm B}$) 
obtained by solving the `Schr\"odinger' equation
\begin{eqnarray}
& & \left[ -\partial_{x'x'} - \partial_{\tau' \tau'} + 
(1-\vphi_{0,{\rm B}}(x',\tau') ) \right] \psi(x',\tau') \nonumber \\
& & \hspace{5mm} = \lambda_i^{(0,{\rm B})} \psi(x',\tau').
\label{Schroedinger}
\end{eqnarray}
If there is a zero-frequency Goldstone mode 
around the bounce solution ($\lambda_i^{({\rm B})} = 0$),
this mode must be replaced by the translational 
mode by using Fadeev-Popov technique.~\cite{Callan77}

{\it Thermal activation regime.} At high temperatures
$T>T_0\sim g^2 \eta^{1/4}$, the bounce solution 
$\vphi_{\rm B}(x',\tau')$
is independent of the imaginary time $\tau'$, and
the ${\rm Im}F$ method reproduces the Kramers-type nucleation rate 
$\Gamma = A \exp(-\Delta U/T)$.~\cite{Kramers40,Haenggi90b}
The energy barrier $\Delta U$ is calculated as
\begin{eqnarray}
\Delta U &=& U[\vphi_{\rm B}(x')] - U[\vphi_0(x')], \\
\frac{U[\vphi(x')]}{(2\eta)^{5/4}}&=& \int_{-\infty}^{\infty} 
{\rm d}x' \left[ \frac{\vphi_{x'}^2}{2} 
+ \frac{\vphi^2}{2} - \frac{\vphi^3}{6}
- \til{\eps} \vphi \delta(x') \right].
\end{eqnarray}
The stationary solutions are obtained
as $\vphi_{\rm B}(x')=\vphi(x';a_1)$ and
$\vphi_0(x')=\vphi(x';a_2)$ respectively, where
\begin{equation}
\vphi(x';a) = \frac{3}{\cosh^2((|x'|+a)/2)}.
\label{CLS}
\end{equation}
The values of $a_1$, $a_2$ ($a_1 < a_2$) depend on
the effective impurity strength $\til{\eps}$ through
$\til{\eps} = 6\alpha (1-\alpha^2)$ with $\alpha
= \tanh(a/2)$. At the current satisfying
$\til{\eps} = \eps (2\eta)^{-3/4} = 4/\sqrt{3}$, 
the stationary solutions
disappear, and the energy barrier $\Delta U$ becomes zero.
Hence, the classical threshold current is 
modified by the impurity from the homogeneous case
$\eta_{\rm c}=0$ as $\eta_{\rm c}=(\sqrt{3}\eps/4)^{4/3}/2$. 

\begin{figure}[tb]
  \epsfxsize=7cm
  \centerline{\epsfbox{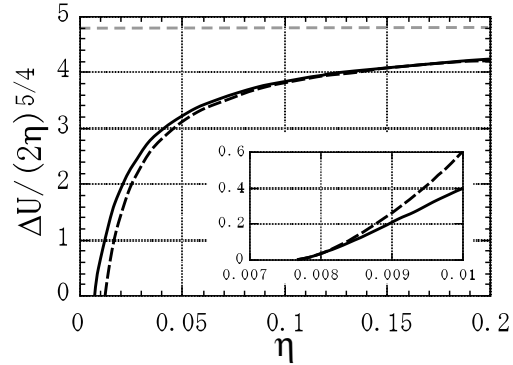}} 
  \vspace{0.1cm}
  \caption{The solid line denotes
  the energy barrier $\Delta U$
  as a function of the bias current $\eta = 1-f$
  obtained numerically 
  for $\eps = 0.1$. The dashed line denotes
  the result (\ref{pert1}) for 
  $\til{\eps} = \eps (2\eta)^{-3/4} \ll 1$, and
  the gray dashed line denotes that of the homogeneous case. 
  The inset shows the behavior of $\Delta U$ near
  the threshold current $\eta = \eta_{\rm c}$
  ($\eta_{\rm c} = 0.076$ for $\eps = 0.1$), 
  and the dashed line in the inset denotes
  the result (\ref{pert2}) for
  $\til{\eps} = 4/\sqrt{3} - \bar{\eps}$ ($\bar{\eps} \ll 1$). }
  \label{FIG1}
\end{figure}

The bias-dependence of the energy barrier $\Delta U$ 
obtained numerically for $\eps = 0.1$ is shown 
by solid lines in Fig.~\ref{FIG1}.
At biases far from the threshold bias ($\eta \gg \eta_{\rm c}$),
the impurity effect is weak, since
the effective impurity strength $\til{\eps}$ is small.
In this case,
the energy barrier $\Delta U$ is proportional to $\eta^{5/4}$ as in
the homogeneous case, and the inhomogeneity
gives only a small correction of order of $\til{\eps} (\ll 1)$ as
\begin{equation}
\Delta U = (2\eta)^{5/4} \left(\frac{24}{5} 
- 3\til{\eps} + {\cal O}({\til{\eps}^2}) \right).
\label{pert1}
\end{equation}
This result is shown by a dashed line in Fig.~\ref{FIG1}.
When $\eta$ approaches the threshold bias 
as $\eta \rightarrow \eta_{\rm c}$, 
the effective impurity strength $\til{\eps}$ approaches
the threshold value $4/\sqrt{3}$, and the bounce solution
is strongly modified. As a result, the bias-dependence of
the energy barrier is modified.
For the strong inhomogeneity
$\til{\eps} = 4/\sqrt{3} - \bar{\eps}$ ($\bar{\eps} \ll 1$), 
the energy barrier $\Delta U$ can be evaluated analytically as
\begin{equation}
\Delta U = \frac{8}{\sqrt{6}} (2\eta_{\rm c})^{5/4} 
\left( \frac{\eta - \eta_{\rm c}}{\eta_{\rm c}} \right)^{3/2}
+ {\cal O} \left(\frac{\eta-\eta_{\rm c}}{\eta_{\rm c}}
\right)^{5/2}.
\label{pert2}
\end{equation}
This result is shown by a dashed line 
in the inset of Fig.~\ref{FIG1}.
Note that the exponent of the bias-dependence is changed from
5/4 (eq.~(\ref{pert1})) to 3/2 (eq.~(\ref{pert2}))
by the strong inhomogeneity.
This crossover about the exponent $B$
at $\eta \sim 2\eta_{\rm c}$ is expected to be detected
experimentally, and is explained by a dimensional
crossover as discussed later.

\begin{figure}[tb]
  \epsfxsize=7cm
  \centerline{\epsfbox{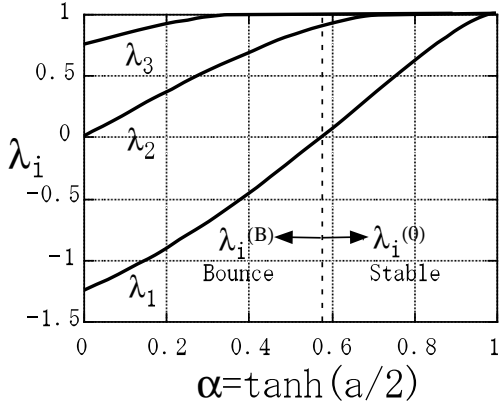}} 
  \vspace{0.1cm}
  \caption{The spectrum of eigenmodes around
  stationary solutions (\ref{CLS}) is shown as a
  function of $\alpha = \tanh(a/2)$.
  For $\alpha < 1/\sqrt{3}$, it
  denotes the spectrum around the bounce solution
  $\vphi_{\rm B}(x')$,
  while for $\alpha > 1/\sqrt{3}$ around the stable solution 
  $\vphi_0(x')$. In addition to these bound states, 
  there exists continuum spectrum at $\lambda \ge 1$.}
  \label{FIG2}
\end{figure}

The prefactor $A$ is also affected by the 
local inhomogeneity through
the change in the spectrum of $\lambda_i^{(0,{\rm B})}$s.
Generally, this change in the spectrum does not affect
the prefactor so strongly as the change of the exponent.
The change of the zero-frequency mode, however,
may produce significant effects on $A$. 
In the homogeneous case, this zero-frequency
mode around the bounce solution $\vphi_{\rm B}(x')$
produces the prefactor proportional to the system size $L$.
In the presence of the inhomogeneity, this mode has a
positive frequency, and modifies the $L$-dependence of $A$.

The spectrum of $\lambda_i^{(0,{\rm B})}$s obtained numerically from 
the Scr\"odinger equation (\ref{Schroedinger})
is shown in Fig.~\ref{FIG2}. The region
$\alpha > 1/\sqrt{3}$ corresponds to the stable solution,
and $\alpha < 1/\sqrt{3}$ to the bounce solution.
The lowest mode $\lambda_1^{({\rm B})}$ denotes the 
variable along the tunneling path. 
The second mode $\lambda_2^{({\rm B})}$
corresponds to the translational mode in the homogeneous case. 
As seen in Fig.~\ref{FIG2}, the frequency of the
second mode remains small for the 
weak inhomogeneity $\til{\eps} \ll 1$
($\alpha \ll 1$). Hence, this mode must be treated
carefully for $\til{\eps} \ll 1$ by the Fadeev-Popov technique. 
Within the first-order perturbation for $\til{\eps}$, 
the contribution of the second mode is treated as
\begin{equation}
\sqrt{\frac{1}{\lambda_2^{(B)}}} \rightarrow
\sqrt{\frac{(2\eta)^{5/4}}{2\pi T} \frac{24}{5}} \int
\! {\rm d}x_0 \exp
\left[- \frac{\Delta U_{\rm imp}/T}{\cosh^2(C x_0/2)} \right],
\end{equation}
where $C=\sqrt{5/24}$. The energy modulation
$\Delta U_{\rm imp}=3\til{\eps} (2\eta)^{5/4} =
3 \eps (2\eta)^{1/2}$ by the inhomogeneity determines 
the $L$-dependence. The prefactor $A$ is 
proportional to $L$ for $\Delta U_{\rm imp} \ll T$ 
($\eta \ll \eta_{\rm cr}$) as in the homogeneous case.
However, $A$ becomes independent of
$L$ for $\Delta U_{\rm imp} \gg T$ ($\eta \gg \eta_{\rm cr}$).
The crossover bias is estimated as $\eta_{\rm cr} 
\sim (T \ln L/\eps)^2$. Note that this crossover at 
$\eta \sim \eta_{\rm cr}$
about $A$ is independent of the crossover near 
$\eta \sim 2 \eta_{\rm c}$ about the exponent $B$.

For the strong inhomogeneity $\til{\eps} = 4/\sqrt{3} - \bar{\eps}$
($\bar{\eps} \ll 1$), the frequency of the translational mode is
so large that its contribution to the prefactor becomes independent
of $L$. Only the frequencies of the lowest modes,
$\lambda_{1}^{({\rm B})}$ and $\lambda_{1}^{({\rm B})}$,
become small compared with the characteristic
frequency of this system ($\lambda = 1$) and the other frequencies of
the eigenmodes. 
Hence, only the lowest mode is relevant to the nucleation process.
To clarify this situation, the field $\vphi(x')$
is truncated to a one-variable problem
\begin{equation}
\vphi(x') = \vphi(x';a_{\rm c}) + C_1(\tau') \psi_1(x';a_{\rm c}),
\end{equation}
where $\vphi(x';a)$ is given in (\ref{CLS}), and
$\psi_1(x';a_{\rm c})=C' \sinh((|x'|+a_{\rm c})/2)
/\cosh^3((|x'|+a_{\rm c})/2)$ is the local deformation mode
which becomes the zero-frequency mode at $a = a_{\rm c}$.
The normalization constant $C'$ is obtained from
$\int {\rm d}x' |\psi_1|^2 = 1$ as $(135/(36-8\sqrt{3}))^{1/2}$,
and the critical value $a_{\rm c}$ is determined by
$\tanh(a_{\rm c}/2) = 1/\sqrt{3}$. The coefficient $C_1$ has dynamics in
the $\tau$-direction under the potential
\begin{equation}
U[\vphi(x')] = {\rm const.} + \frac{2\sqrt{3}}{9} \bar{\eps}
(C' C_1) - \frac{4}{243} (C' C_1)^3.
\label{Ueff}
\end{equation}
Thus, the nucleation is described by the one-variable
potential produced by local deformation of the field.
In this situation, the prefactor $A$ cannot have
the $L$-dependence, because the system size $L$ is irrelevant
to the nucleation process caused by the local deformation
of the field $\vphi(x')$.
Note that this potential form (\ref{Ueff}) reproduces the expression
of the energy barrier (\ref{pert2}) for the strong inhomogeneity.

\begin{figure}[tb]
  \epsfxsize=7cm
  \centerline{\epsfbox{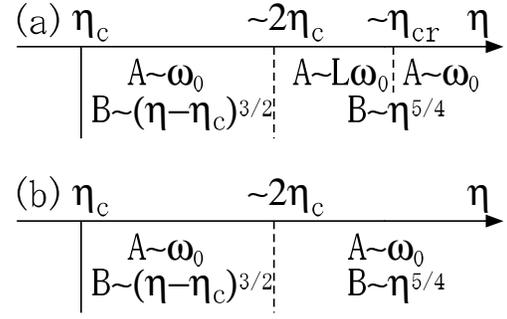}} 
  \vspace{0.1cm}
  \caption{The qualitative behaviors of the nucleation rate
  $\Gamma = A e^{-B}$ (a) for $\eta_{\rm cr} \gg \eta_{\rm c}$ and (b)
  for $\eta_{\rm cr} \ll \eta_{\rm c}$. The characteristic
  attempt frequency is denoted with $\omega_0$.}
  \label{FIG3}
\end{figure}

The result is summarized in Fig.~\ref{FIG3}
for two possible cases. In 
the case of $\eta_{\rm cr} \gg \eta_{\rm c}$,
the prefactor is proportional to $L$ for 
the region $2\eta_{\rm c} < \eta < \eta_{\rm cr}$, 
and it is independent of $L$ for the other current region.
In the case of $\eta_{\rm cr} \ll \eta_{\rm c}$, the
prefactor is independent of $L$ for any $\eta$.
The crossover at $\eta \sim \eta_{\rm cr}$ 
disappears because $\eta_{\rm cr}$ estimated for
$\til{\eps} \ll 1$ is not valid for the strong inhomogeneity
$\eta < 2\eta_{\rm c}$ ($\til{\eps} \sim 1$),
where the prefactor $A$ is independent of $L$ for any $\eta$.

{\it Quantum tunneling regime.} At low temperatures
$T<T_0\sim g^2 \eta^{1/4}$, nucleation due to quantum tunneling 
is dominant. In this regime, the stationary
solutions of the action (\ref{SE2}) must be calculated by
solving the 1+1-dimensional classical field equation.
However, as shown in the thermal activation
regime, the features of the bias-dependence
of the nucleation rate can be discussed by the perturbational
treatment. Hence, in this Letter, only the limiting cases
are discussed to clarify the 
bias-dependence of the nucleation rates.

For the weak inhomogeneity $\til{\eps} \ll 1$, 
the exponent $B$ is obtained within the first-order
perturbation for $\til{\eps}$ as
\begin{equation}
B = \frac{2\eta}{g^2} \left[ s_0 + s_1 \til{\eps} + {\cal O}
(\til{\eps}^2) \right],
\end{equation}
where $s_0 = 31.00$ and $s_1 = 16.43$. In this region, the exponent
$B$ is proportional to $\eta^1$ as in the homogeneous case, and
the inhomogeneity effect only appears as a small correction. 
For the strong inhomogeneity $\til{\eps} = 4/\sqrt{3} - \bar{\eps}$
($\bar{\eps} \ll 1$), the system can be
truncated to the one-variable problem under the potential (\ref{Ueff}).
In this region, the exponent $B$ is obtained as
\begin{equation}
B \simeq \frac{15.8}{g^2} (\eta - \eta_{\rm c})^{5/4}.
\end{equation}
Thus, the exponent of the bias-dependence is changed from
$1$ to $5/4$ by the strong inhomogeneity
at the crossover current $\eta \sim 2\eta_{\rm c}$.

The system-size dependence of the prefactor $A$ can be studied
by a discussion parallel to that of the thermal activation regime.
In the homogeneous case $\til{\eps}=0$, there are two zero-frequency modes
related to the spatial (temporal) translational symmetry of
the bounce soliton $\vphi_{\rm B}^{(0)} (x',\tau')$.
In the weakly-inhomogeneous case $\til{\eps} \ll 1$,
the frequency of the temporal translational mode 
remains zero, while the frequency of the spatial translational mode
is lifted. The Fadeev-Popov technique is applied
to this spatial mode as
\begin{equation}
\sqrt{\frac{1}{\lambda_2^{(B)}}} \rightarrow {\rm const.}
\int {\rm d}x_0 \exp
\left[ - \frac{2\eta \til{\eps}}{g^2} f(x_0) \right].
\label{QTF}
\end{equation}
Here, the function $f(x_0) = \int {\rm d}\tau' 
\vphi_{\rm B}^{(0)}(x_0,\tau')$ behaves as $f(x_0)
\rightarrow 0$ for $|x_0| \rightarrow \infty$, and
has a maximum value $4.784$ at $x_0 = 0$. It is found 
from (\ref{QTF}) that
the factor $\til{\eps} \eta/g^2$ determines the 
system-size dependence of the prefactor $A$. For
$\til{\eps}\eta/g^2 \ll 1$ ($\eta \ll \eta_{\rm cr}$),
the prefactor is proportional to $L$, while for
$\til{\eps}\eta/g^2 \gg 1$ ($\eta \gg \eta_{\rm cr}$),
the prefactor becomes independent of $L$. The 
crossover bias is estimated as $\eta_{\rm cr} \sim
(g^2 \ln L/\eps)^4$.

{\it Discussion.} 
The bias-dependence of 
the exponent $B$ is summarized in Table~\ref{TABLE1},
where the results for small Josephson junctions (SJJs) are also
shown for comparison. At high temperatures,
the exponent $B$ behaves for SJJs as $B \propto \eta^{3/2}$,
and for homogeneous LJJs as $B \propto \eta^{5/4}$.
This difference between SJJs and LJJs comes from
the character of the bounce solution $\vphi_{\rm B}$:
the bounce solution has spatial dependence 
in the $x'$-direction for LJJs,
while it is independent of $x'$ for SJJs. In other words, there is
a spatial dimensional crossover between SJJs and LJJs.
Then, the bias-dependence of $B$ is expressed as 
$B \propto \eta^{3/2 - d/4}$ by a spatial dimension
$d$ which takes $d = 1$ for LJJs and $d = 0$ for SJJs.
There also exists a `temporal' dimensional crossover
between the thermal activation regime and the quantum tunneling
regime: the bounce solution $\vphi_{\rm B}$
has a temporal dependence in the $\tau'$-direction
at low temperatures, while it is independent of $\tau'$
at high temperatures.
As a result, the bias-dependence of $B$ at low temperatures
is expressed
as $B \propto \eta^{3/2 - (d+1)/4}$, where $(d+1)$ denotes
the total dimension of the system including the $\tau$-direction.

The inhomogeneity effects on the bias-dependence of $B$ 
can be understood in terms of a {\it dimensional crossover}.
The inhomogeneity modifies
both the bounce solution $\vphi_{\rm B}(x',\tau')$ and
the stable solution $\vphi_{\rm 0}(x',\tau')$. Although these
two solutions have a spatial dependence, the difference
between these solutions becomes small in the presence
of the strong inhomogeneity. As a result, the nucleation
process is effectively
described by local deformation of the field, and
the spatial dimension $d$ is effectively reduced 
from $d=1$ to $d=0$ by the strong inhomogeneity.
Therefore, in strongly-inhomogeneous LJJs,
the exponent of $\eta$ in $B$ is the same as that of SJJs.
(See Table~\ref{TABLE1}.)

\begin{table}[tb]
\begin{tabular}{lll}
\hline
& high-$T$ & low-$T$ \\
\hline
homogeneous LJJ &
$B \propto \eta^{\frac32-\frac14}$/T & 
$B \propto \eta^{\frac32-\frac14-\frac14}/g^2$ \\
\hline
SJJ &
$B \propto \eta^{\frac32}/T$ &
$B \propto \eta^{\frac32 - \frac14}/g^2$ \\
\hline
inhomogeneous LJJ &
$B \propto (\eta-\eta_{\rm c})^{\frac32}/T$ &
$B \propto (\eta-\eta_{\rm c})^{\frac32 - \frac14}/g^2$ \\
\hline
\end{tabular}
\caption{The bias-dependence of the exponent $B$ in three
systems: homogeneous (or weakly-inhomogeneous)
long Josephson junction (LJJ),
small Josephson junction (SJJ),
and LJJ with the strong inhomogeneity $\til{\eps} \sim 1$.}
\label{TABLE1}
\end{table}


In this Letter, soliton-pair nucleation rates
$\Gamma = A \exp (-B)$ have been studied 
for highly-biased sine-Gordon (SG) systems with a local
inhomogeneity. It is found that
the effective inhomogeneity strength
$\til{\eps} \sim \eps (1-f)^{-3/4}$ controls 
the bias-dependence of the exponent $B$
which is typically written as 
$B \propto (f_{\rm c}-f)^{3/2 - d/4}$ in the
thermal activation regime and as 
$B \propto (f_{\rm c}-f)^{3/2 - (d+1)/4}$ in the
quantum tunneling regime, where $f_{\rm c}$ and $f$ are the
classical threshold bias and the external bias,
respectively. The spatial
dimension $d$ takes $d=1$ for the homogeneous or
weakly-inhomogeneous one-dimensional SG systems ($\til{\eps} \ll 1$),
while it is reduced to $d=0$ 
for strongly-inhomogeneous SG systems ($\til{\eps} \sim 1$).
This change in $B$ is expected to be 
detected experimentally, {\it e.g.} in LJJs.~\cite{Davidson85,Ustinov92} 
This phenomena would be available
to evaluate inhomogeneity strength in real experimental systems. 
It is also found that there
exists a different crossover bias $\eta_{\rm cr}$ at which
the prefactor becomes independent of the system size $L$.
These results are
universal for systems where the nucleation process is 
most dominant. Details of the relevance to actual experiments
will be presented elsewhere.

The author thanks M. Yumoto for stimulating discussion.
The author is supported by Research Fellowship of 
Japan Society for the Promotion of Science 
for Young Scientists. This study was supported by a Grant-in-Aid
for Scientific Research from the Japanese Ministry of Education,
Science, Sports and Culture.

\end{document}